# FROM MICROSCOPIC FLUCTUATIONS TO SUSCEPTIBILITY SPECTRA: SINGLE-MOLECULE RELAXATION IN GLASSY MEDIA

Siyang Wang[#], Jaladhar Mahato[#], Laura J. Kaufman*
Department of Chemistry, Columbia University, New York, New York 10027, USA

[#]These authors contributed equally

*To whom correspondence should be addressed: kaufman@chem.columbia.edu

## Abstract

Single-molecule (SM) rotational dynamics of fluorescent probes in polystyrene near the glass transition temperature ($T_g$) are investigated over long times to reconstruct susceptibility spectra. The loss spectrum, commonly recorded using external field-driven (frequency-domain) spectroscopy, such as dielectric spectroscopy, is reconstructed from purely thermal SM rotational fluctuations. The results reproduce time-temperature superposition typically seen in dielectric spectroscopy for materials near $T_g$ and show that the ensemble spectrum is comprised of individual molecular responses to distinct environments.

## Introduction

Despite decades of intensive study, the microscopic origins of the anomalous dynamics characteristic of glassy systems remain incompletely understood. One of the most striking and widely studied features of glass-forming materials is dynamic heterogeneity, in which dynamics exhibit pronounced variations in both space and time. Numerous theoretical and experimental studies of molecular and polymeric glass formers near the glass transition temperature ($T_g$) have shown that different regions of a supercooled liquid can exhibit relaxation rates spanning several orders of magnitude.[1–4]

Experimental efforts to characterize dynamic heterogeneity have employed a wide range of approaches in both frequency and time domains. Frequency-domain techniques, most notably dielectric spectroscopy, involve the application of sinusoidal external perturbations and measurement of linear response to yield dielectric loss spectra.[5–7] These spectra reveal pronounced non-Debye relaxation, secondary relaxation processes, and often asymmetric dielectric loss peaks that are typically described using phenomenological formalisms.[8–11] Frequency-domain techniques are also well suited for probing aging dynamics, since relaxation spectra can be acquired on experimentally accessible timescales, allowing time-dependent changes associated with aging below $T_g$ to be directly monitored.[11–13] However, dielectric spectroscopy has several inherent limitations. Interpretation of dielectric spectra can be complicated by contributions from sample geometry, conductivity, and interfacial polarization effects, and the technique fundamentally relies on the presence of molecular dipoles.[14–16] Consequently, systems lacking sufficient dipolar response are difficult to probe directly. Although introducing dipolar probes or dopants can enable such measurements, the probe can perturb the local dynamics and alter the relaxation behavior, while the measured response reflects a combination of probe and host contributions, complicating interpretation of the intrinsic dynamic heterogeneity underlying broadened loss spectra.[17,18]

While dielectric spectroscopy is the most common approach to characterizing dipolar relaxations in condensed phase systems, complementary insight has been obtained from time-domain experiments, such as nuclear magnetic resonance (NMR), which typically involves the application of an external field to align dipoles and monitor their return to equilibrium.[19,20] Analogously, rotational dynamics have also been probed using fluorescent probe-based time-domain techniques. These include orientation recovery measurements in fluorescence recovery after photobleaching and correlation-function based analysis of thermal fluctuations via polarization-resolved fluorescence correlation spectroscopy (FCS) and imaging-based FCS.[21–23] In glassy systems, the resulting relaxation from such time-domain experiments often manifests as a stretched exponential characterized by sub-unity stretching exponents $\beta < 1$.[19,20,24,25] While these measurements provide a well-established reference for glassy dynamics, they inherently reflect ensemble-averaged responses, in which contributions from spatially and temporally heterogeneous dynamics are superimposed.[26]

Single-molecule (SM) approaches offer a complementary perspective by eliminating ensemble averaging and directly resolving the dynamics of individual probe molecules embedded in glass-forming hosts.[26] In typical time-resolved SM measurements, trajectories of the linear dichroism (LD) of individual fluorescent probe molecules are analyzed to characterize rotational dynamics of the host system.[27–31] So long as probes are chosen carefully and present at very low concentrations, these measurements yield time correlation functions that encode local dynamics of the host small molecule supercooled liquids[30] or segmental dynamics in polymer matrices,[31] thereby revealing heterogeneity at the molecular scale and providing direct access to the distribution of local dynamics underlying bulk behavior. Nevertheless, the predominantly time-domain nature of SM measurements, in contrast to the frequency-domain character of bulk dielectric spectroscopy, complicates direct quantitative comparison between these two approaches.

Many experimental measurements of glassy relaxation in both the time and frequency domains rely on field-driven response protocols, in which relaxation dynamics are inferred either from the decay to equilibrium following an externally applied perturbation or from the linear response to a weak oscillatory field. In contrast, SM rotational measurements yield time domain information solely from spontaneous thermal fluctuations without any externally applied field. The conceptual link between these perspectives is provided by Onsager's regression hypothesis[32,33] and Kubo's formalism of the fluctuation–dissipation theorem (FDT)[34,35], which relate equilibrium correlation functions to the causal linear response and, via Fourier transformation, to the frequency-dependent susceptibilities measured in bulk spectroscopy. Despite this formal correspondence, it has not been experimentally established whether susceptibility spectra can be quantitatively reconstructed from passive SM rotational fluctuations of probe molecules, nor whether key findings on glassy dynamics obtained through field-driven frequency domain experiments, such as time–temperature superposition (TTS), are also seen when individual molecular trajectories are evaluated. In close analogy to passive microrheology, where single-particle translational fluctuations are mapped onto viscoelastic moduli, mapping SM rotational fluctuations to microscopic susceptibility would provide a direct route for connecting molecular length scale heterogeneity to macroscopic response.[36,37]

Here, we develop and implement such a mapping by reconstructing dielectric-like susceptibility spectra from polarization-resolved SM measurements of probe molecules embedded in polystyrene. From $LD$ trajectories of individual probes, we compute rotational autocorrelation functions and use the Onsager-Kubo/FDT framework to obtain a corresponding time-domain response function, which is then Fourier transformed to yield microscopic susceptibility spectra. The resulting quasi-ensemble spectra, obtained by combining many SM measurements, enables direct comparison to bulk dielectric loss data. This approach demonstrates (1) how the ensemble dielectric relaxation response emerges from the superposition of heterogeneous SM thermal fluctuations and (2) that TTS remains evident when evaluated through passive SM dynamics over a narrow temperature range near $T_\mathrm{g}$.

## Theory: Microscopic Susceptibility from Equilibrium Orientation Fluctuations

The method reported here uses SM imaging to measure spontaneous thermal orientation fluctuations of individual fluorescent probe molecules without external perturbation (Figure 1). This passive approach differs from many conventional bulk methods, which infer susceptibility either from relaxation following an imposed perturbation (time domain) or from the steady-state response to a sinusoidal driving field (frequency domain).[7,20] Our goal is to assess whether susceptibility spectra can be reconstructed directly from spontaneous thermal fluctuations captured in SM trajectories. In our experiments, the dynamic observable is the linear dichroism trajectory $LD(t)$ of SMs, constructed from fluorescence intensities in two orthogonal polarization channels. We define $A(t) \equiv LD(t)$ and the corresponding fluctuation $\delta A(t) = A(t) - \langle A \rangle$, where $\langle \cdots \rangle$ denotes an ensemble average; in practice, this average is estimated from stationary SM trajectories (see Methods). Orientational dynamics are then characterized by the normalized autocorrelation function,

$$C(t) = \frac{\langle \delta A(t)\delta A(0)\rangle}{\langle (\delta A)^2 \rangle}. \quad (1)$$

Although $LD(t)$ does not directly report angular orientation, its equilibrium fluctuation reports rank-2 orientational dynamics under high–NA detection.[27,38–40]

The conceptual basis connecting dielectric spectroscopy and SM LD measurements is Onsager's regression hypothesis together with Kubo's formulation of FDT.[32–35] Onsager's regression motivates the identification of relaxation dynamics with the decay of the equilibrium correlation function, while Kubo's FDT provides the quantitative connection between relaxation function and linear response. Accordingly, the equilibrium autocorrelation function $C(t)$ encodes the same relaxation that would govern the return to equilibrium following a weak perturbation conjugate to $A$.

To make this connection explicit, for a generalized force $f(t)$ conjugate to $A$, linear response gives

$$\delta\langle A(t)\rangle = \int_{-\infty}^{t} \chi(t-t') f(t')\, dt', \qquad (2)$$

where $\chi(t)$ is causal. In our experiment $f(t) = 0$ and $\chi(t)$ is obtained directly from equilibrium fluctuations. In the classical limit relevant here, Kubo's FDT gives

$$\chi(t) = -\frac{1}{k_B T}\Theta(t)\frac{d}{dt}\langle \delta A(t)\delta A(0)\rangle, \qquad (3)$$

with $\Theta(t)$ the Heaviside step function enforcing causality.[35,41] The experimentally relevant frequency-domain quantity is then obtained by Fourier transforming the causal response function,

$$\chi(\omega) = \int_{0}^{\infty} \chi(t) e^{i\omega t}\, dt, \qquad (4)$$

where its imaginary part, $\chi''(\omega)$, is the dissipative (loss) component directly analogous to bulk dielectric loss spectra. In the following sections, we apply this Onsager-Kubo construction to the measured $LD$ trajectories to reconstruct microscopic susceptibility spectra from spontaneous SM dynamics (Figure 1), enabling quantitative comparison with bulk dielectric loss spectra.

With the microscopic susceptibility spectrum as a reference, we extend the same Onsager-Kubo framework from the ensemble level to the level of individual probes. The justification for proceeding molecule-by-molecule is that Onsager's regression and Kubo's FDT relate equilibrium fluctuations to the corresponding causal response for a chosen observable, independent of whether the underlying statistics are obtained from an ensemble of molecules or from a stationary time series of one molecule. Thus, each probe's spontaneous $LD$ fluctuations can be mapped, within the same formalism, to a molecule-specific response function and susceptibility spectrum, enabling direct comparison between the distribution of SM susceptibility spectra and the microscopic susceptibility reference.

## Materials and Methods

### Sample Preparation

In this study, atactic polystyrene (PS, $M_w$ = 168 kg/mol, PDI = 1.05; Polymer Source) was dissolved in toluene and purified by precipitation into hexane three times. The purified polymer was redissolved in toluene to obtain a ~3.5 wt % PS solution. This solution was photobleached for one week using a home-built, high-power light-emitting diode (LED) setup to produce a fluorescence-free stock solution.[42] The fluorescent probe used in this study, N,N'-dipentyl-3,4,9,10-perylenedicarboximide (pPDI; Sigma-Aldrich), was dissolved in toluene and subsequently diluted into an aliquot of the PS stock solution to yield a probe concentration on the order of $10^{-11}$ M. The final solution was spin-coated (Laurell, WS-650MZ-23NPP) at 3000 rpm onto 7 mm × 7 mm silicon wafers (University Wafer) that had been cleaned with piranha solution ($H_2SO_4$:$H_2O_2$ = 2:1). The resulting films were at least 200 nm thick as measured by ellipsometry (J.A. Woollam, alpha-SE), ensuring that the observed dynamics are dominated by bulk behavior.[43] The sample was placed into a home-built microscopy cryostat using vacuum grease (Braycote, 601EF) and annealed at approximately 390 K under ~ 4.0 mTorr vacuum for at least 24 hours to stabilize chamber pressure and remove residual toluene solvent (Figure 2).

### Imaging

Single-molecule data was collected on a home-built wide-field microscope equipped with the cryostat as shown in Figure 2. Specifically, excitation light from a frequency-doubled Nd: Vanadate continuous-wave diode laser at 532 nm (CrystaLaser, CL532-200-LO) was coupled into a multimode fiber (Thorlabs, FG200UCC) that was mechanically shaken with a buzzer at a constant frequency and amplitude to generate a spatially homogeneous, randomly polarized excitation field. The light was subsequently focused onto the back focal plane of an objective lens (Olympus, M Plan Apochromat MPLAPON, 100, NA = 0.95, WD = 0.3 mm). Fluorescence was collected by the same objective in the epi-direction, and directed through a dichroic mirror (Semrock, LPD02-532RU), a long-pass filter (Semrock, LP03-532RE-25), and a bandpass filter (Semrock, FF01-571/72-25) to suppress background noise and excitation leakage. A Wollaston prism (Karl Lambrecht) then split the image into two orthogonally polarized channels, which were simultaneously imaged onto an electron multiplying CCD camera (EMCCD; Andor, iXON DU897).

The glass transition temperature was 374.3 K as determined by differential scanning calorimetry.[44] For pPDI/PS SM measurements, data were collected sequentially at four temperatures (380.9, 377.6, 375.8, and 373.0 K). At each temperature, three movies were recorded. The excitation power measured at the back focal plane of the objective was ~10-25 mW, corresponding to an estimated power density of ~100-250 W/cm² at the sample. Frame rates were selected to yield ~50 frames per median relaxation time at each temperature; from highest to lowest temperature, the times between frames were 0.02 s, 0.1 s, 0.4 s, and 4.6 s, respectively. Movies with time between frames of < 0.2 s were collected continuously, while movies with longer time between frames were collected at 0.2 s exposure time, with illumination shuttered between frames to limit photobleaching. The number of frames collected corresponds to trajectory lengths sufficiently long to avoid statistical effects on extracted rotational correlation times associated with short trajectories.[30,45,46] The chosen frame rate and trajectory lengths at each temperature ensure adequate sampling for transformation to the frequency-domain (see Supplementary 1 and 2). The relatively high excitation power densities required for SM measurements can produce local laser-induced heating, leading to nominal deviations between the actual sample temperature and the set temperature. The sample temperature was estimated using the known temperature dependence of pPDI in PS168k.[31] Specifically, a laser-heating correction of 0.0148 K/mW was applied. An additional constant temperature correction relative to the temperature controller reading was also applied.

### Time-Domain Analysis

From the two orthogonally polarized images acquired on the EMCCD at each time point, single molecules were identified and tracked using the ImageJ Particle Tracker plugin after applying drift correction.[47] Short trajectory segments were stitched together using a hierarchical clustering algorithm.[48] For each tracked molecule, $i$, fluorescence intensities in the s and p polarized channels, $I_{s,i}(t)$ and $I_{p,i}(t)$, were obtained at each time point by integrating the signal over a 3×3-pixel region centered on the nearest integer pixel corresponding to the centroid of imaged emitter. Corresponding features in the two polarization channels were paired using the known lateral offset between the split images.

The paired intensities were used to calculate the single molecule *LD*, defined as

$$LD_i(t) = \frac{I_{s,i}(t) - I_{p,i}(t)}{I_{s,i}(t) + I_{p,i}(t)}. \tag{5}$$

Following the notation in the Theory section, we define $A_i(t) \equiv LD_i(t)$ and the corresponding fluctuation,

$$a_i(t) = A_i(t) - \overline{A_i(t)}, \tag{6}$$

where $\overline{A_i(t)}$, denotes a time average over the trajectory of molecule $i$. For each molecule $i$, the time-averaged normalized autocorrelation function was computed as

$$C_i(t) = \frac{\overline{a_i(t'+t)a_i(t')}}{\overline{a_i^2(t')}}, \tag{7}$$

using the autocorrelation routine available in the Statsmodel library.[49]

Each single-molecule autocorrelation function $C_i(t)$ was fit by least squares (Scipy[50]) to the Kohlrausch-Williams-Watts (KWW) stretched exponential form,

$$C_i(t) = C_i(0) \cdot e^{-(\tau/\tau_{fit,i})^{\beta_i}}. \tag{8}$$

Fitting parameters were constrained to the following ranges: $0.3 < C_i(0) < 2.0$; time between frames (s) < $\tau_{fit,i}$ < movie length (s) / 2; $0.01 < \beta_i < 2.0$. Only trajectories containing at least five data points prior to decay of $C_i(t)$ to 0.1 and exhibiting a coefficient of determination $R^2 > 0.9$ were retained. The rotational correlation time for each molecule was calculated from the fitting parameters as

$$\tau_{c,i} = \frac{\tau_{fit,i}}{\beta_i} \cdot \Gamma\left(\frac{1}{\beta_i}\right). \tag{9}$$

From these trajectories, we constructed a quasi-ensemble (QE) autocorrelation, $C_{QE}(t)$, by averaging $C_i(t)$ over the retained set of single molecule correlation functions, i.e., $C_{QE}(t) = \frac{1}{N}\sum_i C_i(t)$.

Image processing, trajectory reconstruction, and autocorrelation analysis were carried out using a custom Python analysis pipeline as described previously.[48] In addition, the general time-domain framework for analyzing SM dynamics has been established and discussed extensively in prior work.[29,51]

### Frequency-Domain Analysis

Autocorrelation functions were transformed into the frequency domain to construct dielectric-like loss spectra. Following the Onsager-Kubo/FDT framework described in the Theory section, the QE autocorrelation function was mapped onto a causal time-domain response function and then Fourier transformed to obtain the frequency-domain microscopic susceptibility.

To enable direct comparison with bulk dielectric loss measurements, we first constructed the QE autocorrelation function $C_{QE}(t)$ as described above. The corresponding QE response function was then obtained from Kubo's FDT as

$$\chi_{QE}(t) = -\Theta(t)\frac{d}{dt}C_{QE}(t), \qquad (10)$$

where $\Theta(t)$ enforces causality. The QE susceptibility spectrum was calculated by Fourier transforming the causal response function. Because direct fast Fourier transform (FFT) of finite-duration time-domain data introduces truncation artifacts and does not reproduce the experimentally observed broad spectral spectrum, the transformation was instead performed numerically via

$$\chi''_{QE}(\omega) = \int_0^\infty \chi_{QE}(t)\sin(\omega t)\,dt. \qquad (11)$$

Time derivatives were computed by finite differences and integrals were evaluated numerically using Simpson's rule. Angular frequencies $\omega$ were sampled on a logarithmically spaced grid spanning the experimentally accessible frequency window set by the movie duration and camera frame interval. This frequency window covered approximately 3-3.5 decades.

$\chi''_{QE}(\omega)$ was then parameterized by fitting to a Havriliak–Negami (HN) form,[52]

$$\chi''_{QE}(\omega) = \chi_\infty + \Delta\chi[\mathrm{Im}\{(1+i\omega\tau_{HN})^\alpha\}]^{-\gamma}, \quad (12)$$

where $\Delta\chi$ is the relaxation strength, $\alpha$ and $\gamma$ report the symmetric and asymmetric broadening, respectively, and $\chi_\infty$ accounts for the high-frequency baseline. From the fitted HN parameters we calculated the peak timescale,[53]

$$\tau_{peak} = \tau_{HN}\cdot\left[\frac{\sin\left(\frac{\pi\alpha}{2(1+\gamma)}\right)}{\sin\left(\frac{\pi\alpha\gamma}{2(1+\gamma)}\right)}\right]^{1/\alpha}. \qquad (13)$$

With $\chi''_{QE}(\omega)$ as a reference, we then computed SM susceptibilities by applying the same mapping for each molecule using its KWW fit. This enables visualization of each molecule's contribution to the overall $\chi''_{\mathrm{QE}}(\omega)$ spectrum. For each molecule $i$, the measured autocorrelation $C_i(t)$ was first fit to the KWW form (Eq. 8), yielding a smooth representation $C_i^{KWW}(t)$. The causal response function was then obtained from the fitted correlation function,

$$\chi_i(t) = -\Theta(t)\frac{d}{dt}C_i^{KWW}(t), \qquad (14)$$

and the corresponding local spectrum was calculated as

$$\chi''_i(\omega) = \int_0^\infty \chi_i(t)\sin(\omega t)\,dt. \qquad (15)$$

This procedure assigns a frequency-domain susceptibility spectrum to each individual probe while reducing numerical noise associated with differentiating finite-length correlation functions. All frequency-domain transformations, averaging procedures, and fitting routines were implemented in a custom Python analysis pipeline.

## Results and Discussion

As proof-of-principle to validate the approach outlined in Methods, we investigated dynamics near the glass transition temperature using extremely dilute, bright, and photostable fluorescent probes embedded in a polymer matrix. Polystyrene, a canonical high glass-transition-temperature polymer, provides an ideal benchmark system for this purpose. Over the past decade, we have used fluorescent probes such as pPDI in polystyrene to elucidate key features of the dynamics of the system near the glass transition.[31,48,54] Our earlier studies revealed that pPDI behaves as an ideal probe for the polystyrene host; it faithfully reports ensemble behavior such as non-Arrhenius temperature dependence, while simultaneously revealing significant heterogeneity evident from the broad distribution of $\tau_{fit}$ and sub-unity $\beta$ values. This benchmark thus provides a firm basis for direct comparison between molecule-resolved observables and bulk relaxation behavior.

To validate consistency with established pPDI/PS behavior, we first compared the time-domain SM dynamical parameters to prior measurements of pPDI rotational dynamics in high–molecular-weight polystyrene. The median SM rotational correlation time, $\tau_{c,med} = median(\tau_{c,i})$, exhibits the expected temperature dependence and agrees with previous reports for comparable PS hosts.[31] Moreover, the distributions of SM relaxation parameters provide clear evidence of dynamic heterogeneity. Specifically, SM autocorrelations are well described by stretched-exponential decays with sub-unity stretching exponents, with median $\beta_i$ = 0.52–0.58 across the temperatures studied (Supplementary 2). Because these results primarily serve to validate data quality and establish consistency with known pPDI/PS behavior, this data is primarily presented in Supplementary Information, and the focus of the main text is on the fluctuation-to-susceptibility construction enabled by and performed on these trajectories.

Figure 3 shows reconstructed microscopic susceptibility spectra from thermal spontaneous SM rotational fluctuations and links time-domain and frequency-domain descriptions of glassy relaxation. From the quasi-ensemble (QE) autocorrelation function $C_{QE}(t)$, constructed by averaging time-averaged SM autocorrelations (Methods), we obtain the corresponding QE causal response function $\chi_{QE}(t)$ using the Onsager-Kubo/FDT mapping (Figure 3a, b). Upon cooling toward $T_{\rm g}$, $\chi_{QE}(t)$ exhibits systematic slowing; its decay reflects the same temperature-dependent relaxation encoded in $C_{\rm QE}(t)$.

The lower panels (Figure 3c, d) show the corresponding quasi-ensemble dielectric-like loss spectra $\chi''_{\rm QE}(\omega)$, obtained by Fourier transforming $\chi_{\rm QE}(t)$. The reconstructed microscopic susceptibility spectra are consistent with those measured by Roland and Casalini;[5] they are markedly broad, spanning decades in frequency with substantial spectral weight and displaying asymmetry – a hallmark of dielectric loss spectra in glass-forming systems.[5–7] The asymmetric peak shifts to lower frequencies with decreasing temperature, consistent with the variation of qualitative features of bulk dielectric loss spectra of polystyrene near $T_{\rm g}$ (Figure 3c). Fits of $\chi''_{\rm QE}(\omega)$ to the Havriliak–Negami (HN) form yield a peak timescale $\tau_{\rm peak}(T)$ that varies smoothly in a non-Arrhenius manner over the measured temperature range and is well described phenomenologically by a Vogel–Fulcher–Tammann (VFT) form. The resulting $\tau_{\rm peak}(T)$ agrees with the median SM rotational correlation time $\tau_c$ obtained from time-domain analysis (Figure 3d).

Having shown that the reconstructed QE susceptibility spectra recover established ensemble trends, we next leverage the SM nature of the data to ask how spectral broadening in such susceptibility spectra emerges. While microscopic susceptibility spectra can, in principle, be reconstructed from fluorescence-based quasi-ensemble rotational measurements, meaningful comparison with the bulk dielectric response requires that the probe's local rotational dynamics faithfully reflect the host relaxation behavior. A distinctive advantage of our SM measurements is that we follow many individual probes for long durations in a molecule-resolved fashion, preserving full spatiotemporal specificity for each trajectory. This high-throughput approach allows us to address a central question: does a bulk-like susceptibility spectrum appear intrinsically broad and asymmetric at the SM level, or do these features arise primarily from averaging over molecules residing in dynamically distinct local environments? To answer this, we use the dataset collected at 375.8 K as a representative example and analyze how QE susceptibility emerges from the distribution of molecule-resolved susceptibilities.

Figure 4 shows the molecule-specific $\chi_i''(\omega)$ spectra at 375.8 K and how they relate to the $\chi''_{\rm QE}(\omega)$ spectrum. Additional temperatures are shown in Supplementary 3. Thin colored curves show $\chi_i''(\omega)$, reconstructed for each molecule by applying the Onsager-Kubo/FDT mapping to its time-domain correlation function (implemented using the molecule-specific KWW fit; Methods). The corresponding QE spectrum at the same temperature is shown as green symbols. All $\chi_i''(\omega)$ are processed using the same analysis pipeline; the QE microscopic susceptibility is plotted on a secondary y-axis for visual clarity (dashed green line) along with the average of the molecule-resolved spectra (red dotted line). A slight difference is observed between the QE spectrum obtained from the QE autocorrelation function and the spectrum obtained by averaging the molecule-resolved susceptibility. This arises because the susceptibility reconstruction is performed using molecule-specific KWW representations, such that averaging the correlation functions before reconstruction is not strictly identical to averaging the reconstructed susceptibilities afterward (Supplementary 4).

Peak positions and line-shapes vary substantially across molecules in $\chi_i''(\omega)$, spanning more than an order of magnitude in $\omega$, directly reflecting dynamic heterogeneity. Molecule-to-molecule differences in the peak position show heterogeneity in the characteristic relaxation times of local environments, while the intrinsic asymmetry and breadth of individual $\chi_i''(\omega)$ spectra reflect non-exponential relaxation within a given trajectory. Thus, molecule-resolved $\chi_i''(\omega)$ spectra enable direct visualization of dynamic heterogeneity in $\chi''_{\mathrm{QE}}(\omega)$, as suggested by the broadening of the bulk dielectric susceptibility measurements relative to the Debye form.

Taken together, the molecule-resolved $\chi_i''(\omega)$ spectra show how the QE response is built from heterogeneous microscopic responses. The QE curve is distinct from any SM trace, consistent with superposition of many distinct local relaxation processes. Figure 4 thus provides a direct visual link between our methodological pipeline and the central physical point raised in the Introduction: bulk-like loss spectra can be quantitatively reconstructed from passive SM fluctuations, and the macroscopic response can be understood as the average over a distribution of molecule-resolved losses whose collective envelope yields $\chi''_{QE}(\omega)$.

Having established the origin of the spectral broadening, this motivates a natural next question: does the QE susceptibility reconstructed from thermal fluctuations of single molecules follow the same temperature-scaling principles as bulk relaxation spectra? Time–temperature superposition (TTS) is a central organizing principle in glass-forming dynamics.[6] Under TTS, the frequency-dependent response retains an approximately invariant spectral shape over a limited temperature window, with temperature entering primarily through a horizontal rescaling of the characteristic timescale (and amplitude). In bulk dielectric spectroscopy, TTS is commonly observed and is often interpreted as evidence that the dominant relaxation mechanism remains unchanged over a limited temperature range near $T_{\mathrm{g}}$, with temperature primarily shifting the spectrum along the frequency axis.[6,55] Testing whether TTS holds for susceptibilities reconstructed from molecule-resolved dynamics therefore provides a stringent benchmark of whether passive SM fluctuations recover the same collective relaxation physics as bulk measurements.

Figure 5 tests TTS for the QE susceptibility spectra reconstructed from SM rotational thermal fluctuations using the Onsager-Kubo/FDT mapping. Each temperature-dependent spectrum $\chi''_{QE}(\omega)$ is rescaled by its peak position and peak amplitude and plotted as $\chi''_{QE}(\omega)/\chi''_{QE,max}(\omega)$.[6,56] Here, if TTS holds, the curves will collapse onto a single master curve. Over the temperature window studied (373.0–380.9 K), the four rescaled $\chi''_{QE}(\omega)$ largely overlap through the peak and into both the low- and high-frequency flanks, indicating an approximately temperature-invariant line-shape in this range and showing that the dominant temperature dependence is captured by a horizontal shift (together with peak normalization). Small deviations are most apparent in the high-frequency wing, where the collapse is imperfect, consistent with the limited accessible frequency band and increased noise at the shortest time lags.

The overlaid fit curves provide compact parameterization of the master spectrum shape. The black curve is a Havriliak–Negami (HN) description of the master lineshape ($\alpha_{HN} \approx 0.89$, $\gamma_{HN} \approx 0.38$), while the orange curve is obtained by numerically transforming the corresponding KWW representation ($\beta_{KWW} \approx 0.49$). The $\alpha_{HN}$ value reflects weak symmetric broadening of the relaxation peak relative to the Debye form, while $\gamma_{HN} \ll 1$ indicates pronounced asymmetry, leading to a stretched high-frequency flank. Both capture the broadened, non-Debye peak shape expected near $T_{\mathrm{g}}$. The ability of a single set of shape parameters to describe the rescaled $\chi''_{QE}(\omega)$ supports TTS for the QE susceptibility over the explored temperature interval.

The results presented here provide microscopic insight into the long-recognized correspondence between SM rotational measurements and bulk dielectric relaxation in glass-forming systems. Prior studies across diverse polymer and molecular glass formers have shown that SM observables can track bulk-like temperature-dependent dynamics;[30,31] however, the correspondence between dielectric measurements that probe an ensemble-averaged linear response and SM measurements accessing molecule-resolved thermal fluctuations has not been demonstrated previously. Building on Onsager's regression hypothesis and Kubo's fluctuation-dissipation formalism, we bridge this gap by reconstructing susceptibility spectra directly from equilibrium SM rotational fluctuations and by examining how a bulk-like response emerges from the distribution of molecule-specific dissipative dynamics. As a proof of principle, we validate this approach using pPDI probes in high–molecular-weight polystyrene near $T_g$, leveraging an improved experimental configuration that enables extended trajectories with enhanced signal-to-background. The resulting quasi-ensemble spectrum - constructed from the collection of molecule-resolved susceptibilities - yields a broadened, asymmetric loss peak consistent with bulk observations without imposing a phenomenological relaxation model *a priori*.

Importantly, this framework also provides insights into the microscopic origin of spectral broadening and its temperature dependence near the glass transition. Variations in microscopic susceptibility peak position across molecules report heterogeneity in the characteristic relaxation times of local environments, while the intrinsic breadth of individual susceptibilities reflects non-exponential relaxation within a given trajectory. The quasi-ensemble spectrum further exhibits time–temperature superposition over the temperature range studied, indicating that despite pronounced heterogeneity, the spectral shape remains approximately invariant, and temperature primarily rescales the characteristic timescale. Together, these findings establish a direct, model-independent connection between molecule-resolved spontaneous thermal fluctuations and ensemble linear response, providing a general framework for interpreting SM measurements in terms of bulk dielectric and viscoelastic relaxation in glass-forming materials. More broadly, this approach provides an alternative route for probing relaxation behavior in systems with insufficient intrinsic dipolar response, where conventional dielectric spectroscopy is difficult to apply directly.

## Conclusions

In summary, we experimentally demonstrate that susceptibility spectra consistent with bulk dielectric measurements can be reconstructed directly from thermal equilibrium single-molecule rotational fluctuations using Onsager's regression hypothesis and Kubo's fluctuation-dissipation framework. By tracking extended single-molecule trajectories near $T_g$, in a well-characterized polystyrene glass former, we show how a bulk-like microscopic susceptibility response emerges from a distribution of molecule-specific dissipative dynamics, providing a microscopic foundation for the correspondence between single-molecule measurements and ensemble-averaged dielectric relaxation.

## Supplementary Material

The supplementary material contains information on how frame rate and trajectory length affect dielectric-like loss spectra, single molecule information obtained from the time domain experiments, transformed single molecule data at three temperatures, and discussion of how the portion of the autocorrelation function fit affects obtained dielectric-like loss spectra.

## Acknowledgements

This work was supported in part by NSF CHE 2246765. SW was supported in part through a Ge Li and Ning Zhao Family Foundation Summer Fellowship. JM thanks Souradeep Gupta for helpful discussion.

## Author Declarations

### Conflict of Interest

The authors have no conflicts to disclose.

### Author Contributions

**Siyang Wang**: Conceptualization (equal); Investigation (lead); Data curation (equal); Formal analysis (lead); Methodology (equal); Software (lead); Validation (equal); Visualization (lead); Writing – original draft (equal); Writing – review & editing (supporting). **Jaladhar Mahato**: Conceptualization (equal); Investigation (supporting); Data curation (supporting); Formal analysis (equal); Methodology (equal); Validation (equal); Visualization (supporting); Writing – original draft (equal); Writing – review & editing (supporting). **Laura J. Kaufman**: Conceptualization (supporting); Project Administration (lead); Methodology (equal); Investigation (supporting); Funding Acquisition (lead); Validation (supporting); Supervision (lead); Writing – review & editing (lead). JM and SW contributed equally to this work.

### DATA AVAILABILITY

The code used in this work is available at https://github.com/Kaufman-Lab-Columbia/SingleMoleculeRelaxationToMicroscopicSusceptibility. The single molecule data that support the findings of this study are available from the corresponding author upon reasonable request.

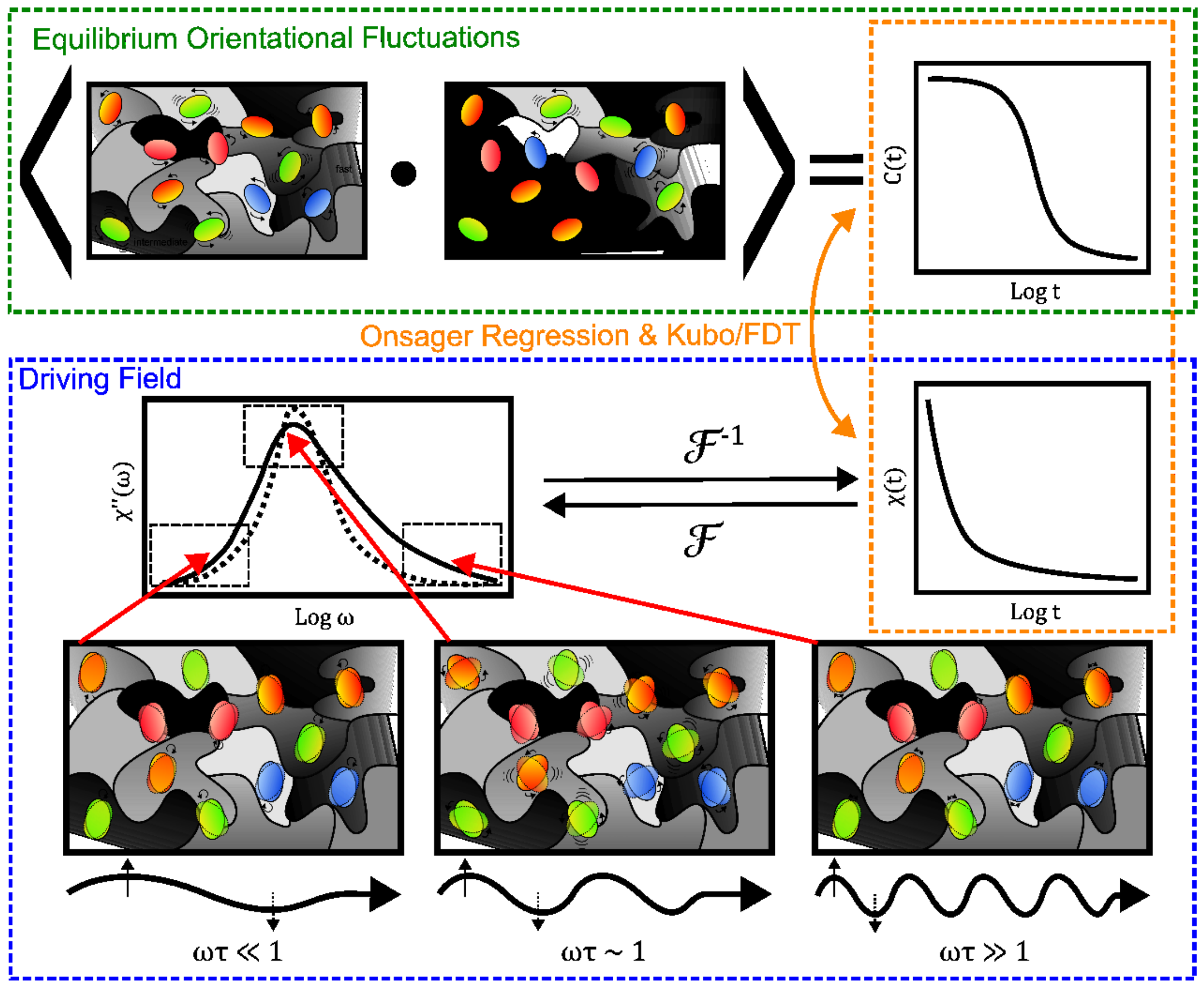

**Figure 1.** Bridging equilibrium single-molecule fluctuations and susceptibility spectra. Top (green box): The depicted single-molecule experiment measures spontaneous equilibrium orientational thermal fluctuations of individual probes. Here, as in all schematics in this figure, the spatiotemporally heterogeneous system is represented in shades of gray, corresponding to domains of differing mobility. Colored ellipses represent probe molecules residing in environments with distinct dynamics. The normalized autocorrelation, C(t), is constructed by correlating probe thermal fluctuations, providing a direct measure of correlation timescales in the absence of any applied perturbation. Bottom (blue box): Conventional bulk dielectric spectroscopy infers the same physics from the steady-state response to a sinusoidal driving field, reporting the frequency-dependent dielectric loss spectrum $\chi''(\omega)$. The solid black line represents non-Debye glassy relaxation, while the dashed black curve indicates an ideal Debye response for comparison. Three frequency regimes of a typical dielectric loss spectrum are highlighted to schematically illustrate the relaxation behavior across different frequency regimes. At low frequencies ($\omega\tau \ll 1$), particles can readily follow the slowly oscillating field; at high frequencies ($\omega\tau \gg 1$), particles are unable to follow the rapidly oscillating field and only exhibit small-amplitude motions, resulting in minimal energy dissipation. Maximum energy loss occurs when the oscillation frequency and relaxation timescale are comparable ($\omega\tau \sim 1$), leading to the characteristic loss peak. Red arrows indicate contributions from each frequency regime to the dielectric loss spectrum, while the black sinusoidal curves represent the applied oscillatory field frequency. Center right (orange box): Onsager's regression and Kubo's fluctuation–dissipation theorem connect these viewpoints by mapping the decay of equilibrium fluctuations onto the causal linear response function, $\chi(t)$, and upon Fourier transformation ($\mathcal{F}$), to the susceptibility spectrum. In this work, we use this Onsager-Kubo/FDT mapping to reconstruct $\chi(t)$ and $\chi''(\omega)$ directly from single-molecule trajectories, enabling quantitative comparison with bulk dielectric loss spectra.

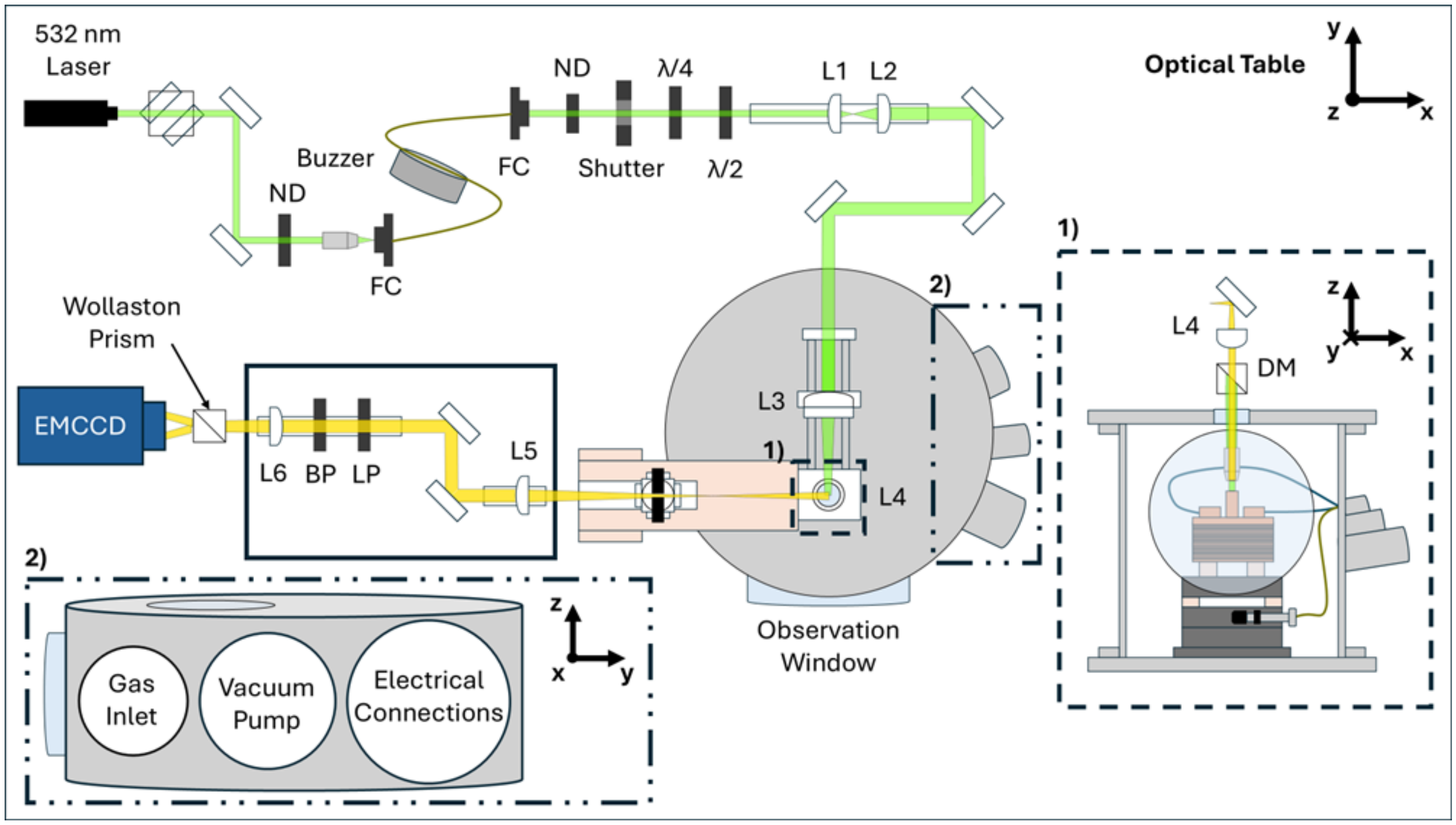

**Figure 2.** Schematic diagram (not to scale) of the home-built wide-field fluorescence microscope configuration used in this study. The system integrates (*i*) a table-top excitation/beam-conditioning module, (*ii*) a vacuum cryostat/sample chamber with optical access (observation window) and in-chamber positioning (piezo stage) (box 1), and ports for gas handling, pumping, and electrical feedthroughs for temperature control (heater and sensor) (box 2), and (*iii*) a dual-polarization detection arm coupled to an EMCCD camera. The sample is mounted inside the cryostat and the emitted fluorescence is collected through the observation window. The emission is then split into two orthogonal polarization channels via a Wollaston prism to generate the single-molecule linear dichroism trajectories used throughout this work. This setup enables the use of a high-NA objective with a short working distance inside the cryostat. Optical elements are abbreviated as follows: ND, neutral density filter; FC, fiber coupling; λ/4 (λ/2), quarter- (half-) waveplate; L, lens; DM, dichroic mirror; BP, bandpass filter; LP, longpass filter.

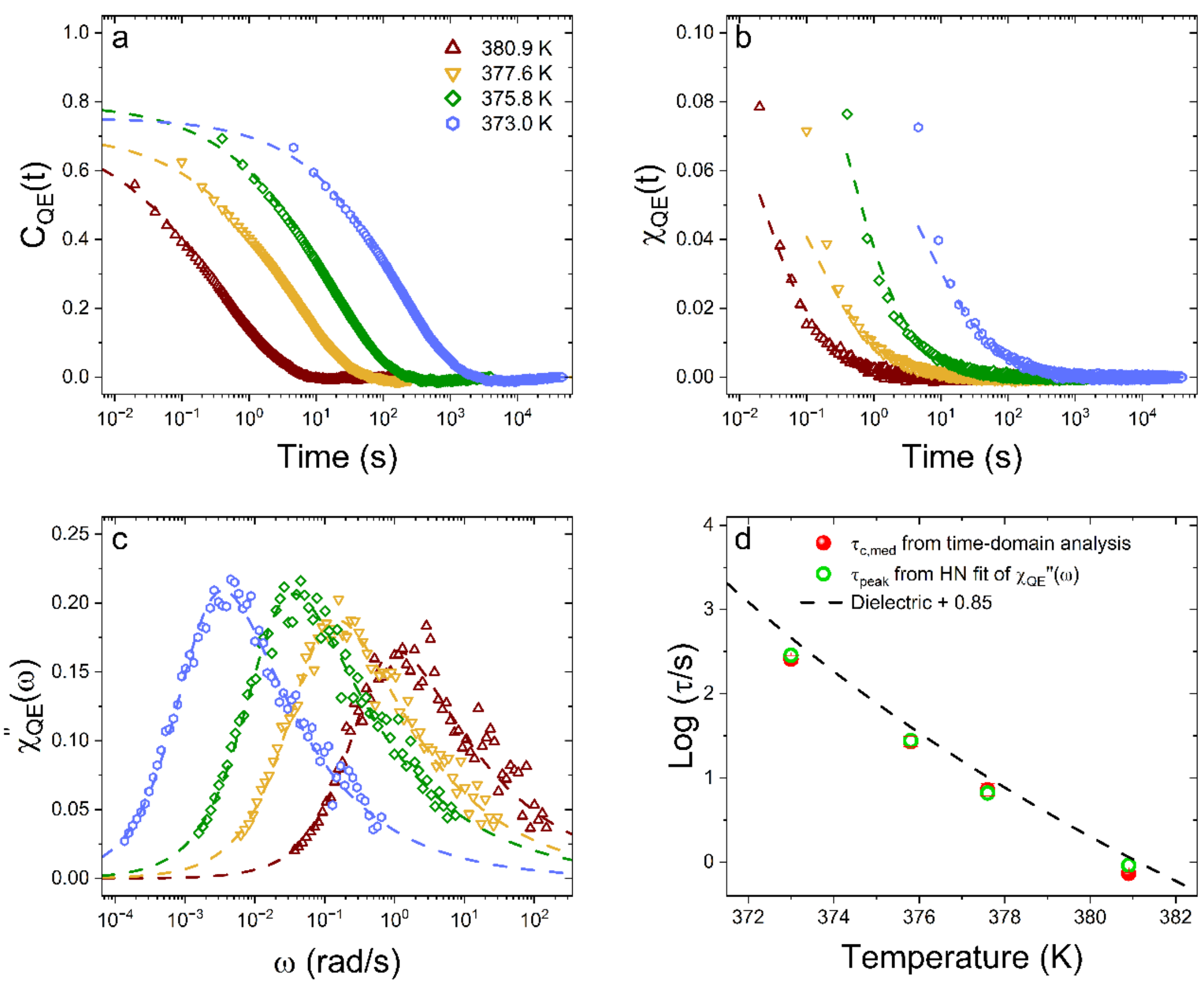

**Figure 3.** Reconstruction of quasi-ensemble microscopic susceptibility spectra from single-molecule rotational thermal fluctuations in pPDI/PS near $T_g$. (a) Quasi-ensemble autocorrelation functions $C_{QE}(t)$ at each measured temperature constructed by averaging single-molecule $LD_i(t)$ autocorrelations; dashed curves show KWW fits, $\beta_{QE}$ = 0.48–0.52. (b) Corresponding causal response functions $\chi_{QE}(t)$ obtained from $C_{QE}(t)$ via the Onsager-Kubo/FDT mapping; dashed curves show response functions implied by the KWW fits in (a). (c) Dielectric-like loss spectra $\chi''_{QE}(\omega)$ obtained by Fourier transforming $\chi_{QE}(t)$; dashed curves show Havriliak–Negami fits, $\alpha$ = 0.89–0.94, $\gamma$ = 0.37–0.41. (d) Temperature-corrected rotational timescales for pPDI in polystyrene at temperatures near $T_g$ from single-molecule time-domain analysis (red, $\tau_{c,med}$) and quasi-ensemble frequency-domain analysis (green, $\tau_{peak}$); the solid curve shows the Vogel–Fulcher–Tammann (VFT) fit for polystyrene reported from dielectric spectroscopy by Roland and Casalini.[5]

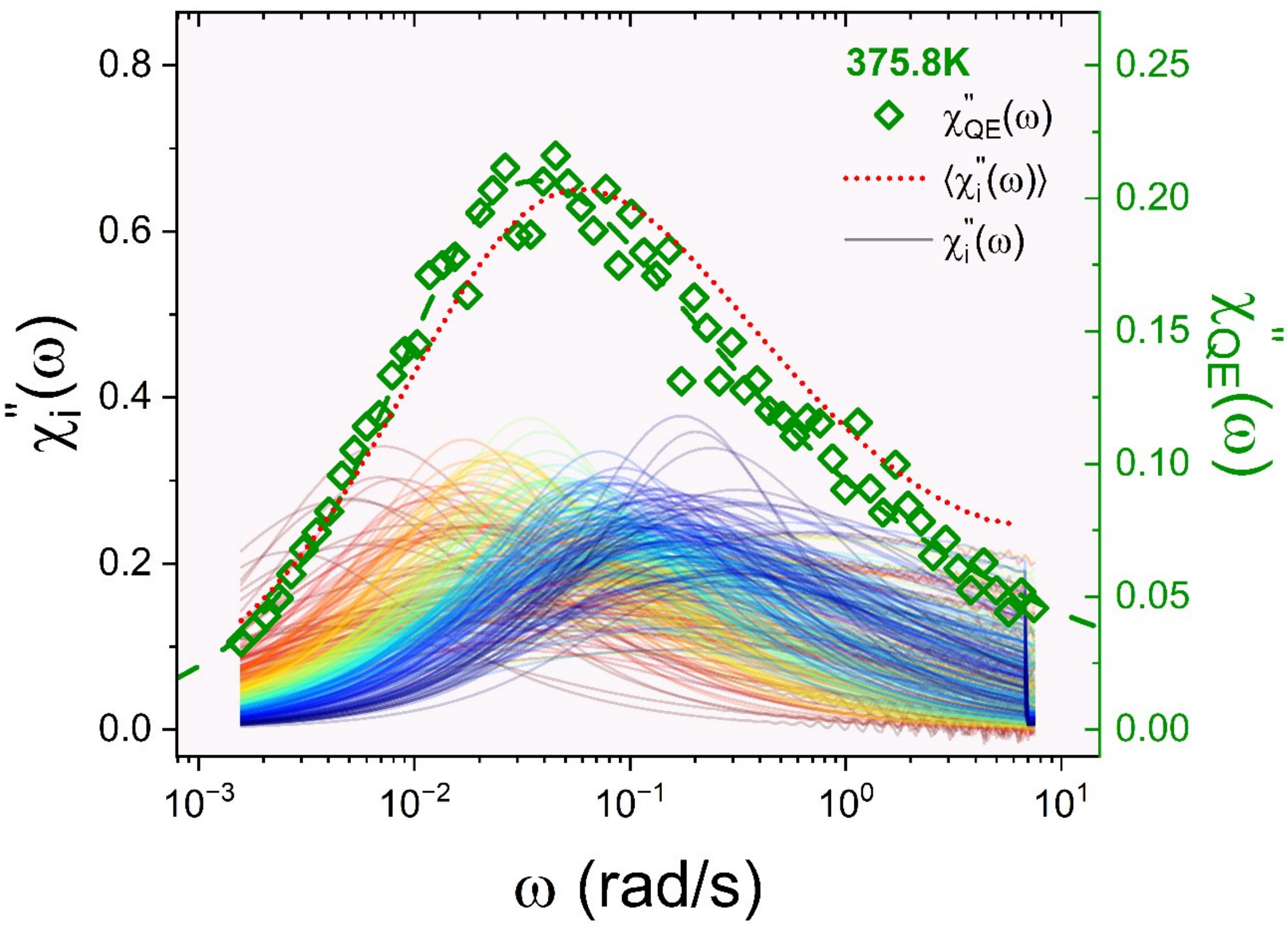

**Figure 4.** Quasi-ensemble (QE) microscopic susceptibility spectrum and constituent single-molecule spectra at 375.8 K. Thin colored curves show molecule-resolved loss spectra $\chi_i''(\omega)$ computed for each probe by applying the Onsager-Kubo/FDT mapping to the KWW-fit of the correlation function for that molecule (see Methods). Green symbols show $\chi_{QE}''(\omega)$ obtained by transforming the raw QE correlation function $C_{QE}(t)$; the green dashed line is the corresponding Havriliak-Negami fit to $\chi_{QE}''(\omega)$. The red dotted line is the average molecule-resolved $\langle\chi_i''(\omega)\rangle$.

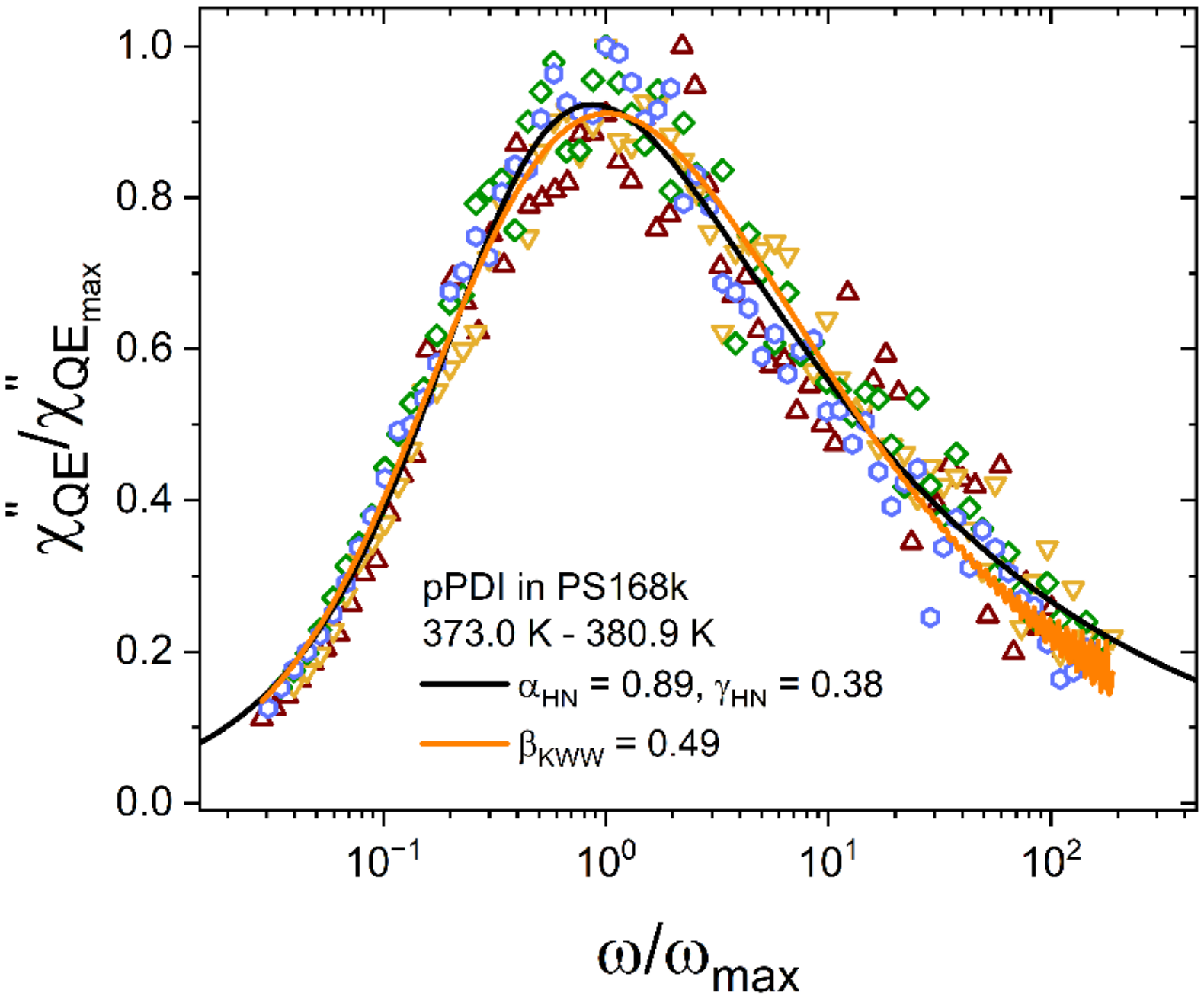

**Figure 5.** Master plot of the quasi-ensemble dielectric-like loss spectra $\chi''_{QE}(\omega)$ of pPDI in PS168k (373.0–380.9 K) also shown in Figure 3c. The solid curves show fits to the master spectrum using a Havriliak–Negami form (black) and a KWW relaxation function evaluated numerically in the frequency domain (orange) (both reporting $R^2 > 0.95$). This collapse demonstrates that time–temperature superposition holds over the investigated temperature range.

## SUPPLEMENTAL ONLINE MATERIAL

# FROM MICROSCOPIC FLUCTUATIONS TO SUSCEPTIBILITY SPECTRA: SINGLE-MOLECULE RELAXATION IN GLASSY MEDIA

Siyang Wang[#], Jaladhar Mahato[#], Laura J. Kaufman*
Department of Chemistry, Columbia University, New York, New York 10027, USA

## Supplementary 1 - Stretched-Exponential Decays and Susceptibility Spectra

## Supplementary Text

Ideal stretched-exponential decays with $\tau_{fit} = 1$ and $\beta = 0.50$ were numerically evaluated using different frame rates and trajectory lengths and subsequently fit to the Havriliak-Negami (HN) form introduced in the Methods section. The results demonstrate that single-molecule data acquisition parameters can influence the apparent shape of the reconstructed susceptibility spectrum, even when all spectra derive from the same underlying relaxation.

Stretched exponential decays numerically evaluated at a lower sampling rate lead to spectra with limited information in the high-frequency wing relative to those performed at a higher sampling rate for a simulation of a given trajectory length (Figure S1a-d). To enhance coverage of the low frequency wing, long trajectories should be collected (Figure S1e, f). To minimize potential misinterpretation of spectral shape and maintain consistency throughout this work, movies at all temperatures were collected using comparable frame rates and trajectory lengths relative to $\tau_{\mathrm{fit}}$. In particular, acquisition conditions were chosen to produce susceptibility spectra similar to Figure S1f, where the low- and high-frequency amplitudes are approximately balanced, analogous to conventional bulk dielectric measurements. Information on single molecule data acquisition parameters is provided in Supplementary 2.

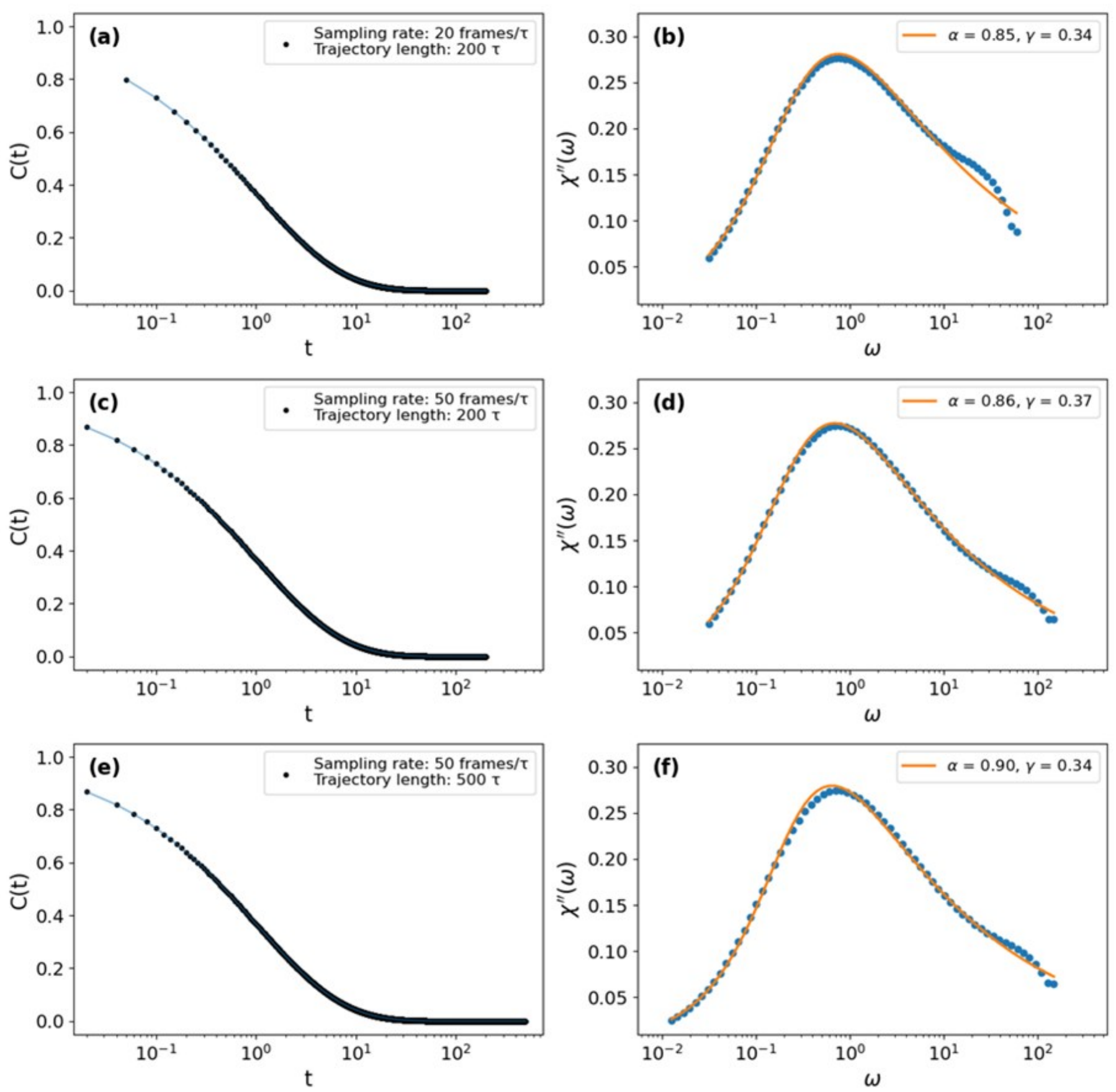

**Figure S1.** (left) Numerically evaluated KWW decays with $\tau_{fit} = 1$ and $\beta = 0.50$ and (right) corresponding susceptibility spectra in which blue circles indicate numerically transformed data, and orange curves show fits to the HN form, illustrating the influence of sampling conditions on extracted spectral shape parameters. (a) KWW curve with frame rate of 20 points per $\tau$ and a trajectory length of 200 $\tau$ yields (b) a susceptibility spectrum best fit to the HN form with $\alpha = 0.85$ and $\gamma = 0.34$. (c) KWW curve with sampling interval of 50 points per $\tau$ and a trajectory length of 200 $\tau$ yields a (d) susceptibility spectrum best fit to the HN form with $\alpha = 0.86$ and $\gamma = 0.37$. (e) KWW curve with sampling interval of 50 points per $\tau$ and a trajectory length of 500 $\tau$ yields (f) a susceptibility spectrum best fit to the HN form with $\alpha = 0.90$ and $\gamma = 0.34$.

**Supplementary 2 – Rotational Dynamics as Obtained from SM Measurements**

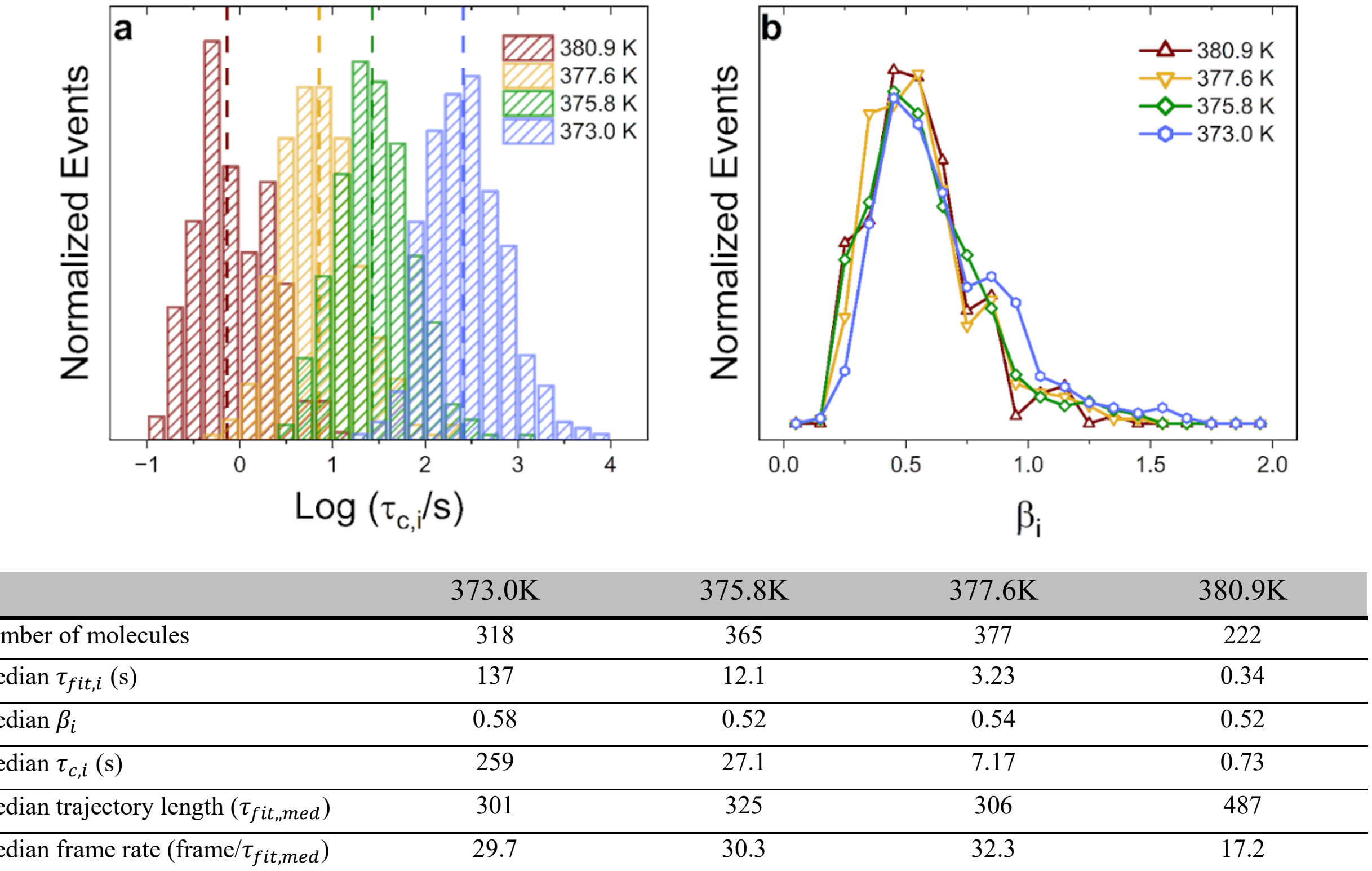

| | 373.0K | 375.8K | 377.6K | 380.9K |
|---|---|---|---|---|
| number of molecules | 318 | 365 | 377 | 222 |
| median $\tau_{fit,i}$ (s) | 137 | 12.1 | 3.23 | 0.34 |
| median $\beta_i$ | 0.58 | 0.52 | 0.54 | 0.52 |
| median $\tau_{c,i}$ (s) | 259 | 27.1 | 7.17 | 0.73 |
| median trajectory length ($\tau_{fit,,med}$) | 301 | 325 | 306 | 487 |
| median frame rate (frame/$\tau_{fit,med}$) | 29.7 | 30.3 | 32.3 | 17.2 |
| $\beta_{QE}$ | 0.52 | 0.48 | 0.48 | 0.50 |

**Figure S2.** (a) Rotational correlation time, $\tau_{c,i}$, and (b) stretching exponent, $\beta_i$, distributions for single molecule rotations at the four temperatures studied. In (a) dashed vertical lines are median values. Median $\beta_i$ values ranged from 0.52–0.58 with no clear trend as a function of temperature. Table below provides key information about the data collected at each temperature: number of molecules, median single molecule $\tau_{fit}$, $\beta$, and $\tau_c$ values, median trajectory length in terms of median single molecule $\tau_{fit}$ value and median frame rate in terms of median $\tau_{fit}$value, as well as the $\beta$ value obtained from the quasi-ensemble correlation functions shown in Figure 3a in the main text.

## Supplementary 3 – Quasi-Ensemble Susceptibility Spectra at Three Temperatures

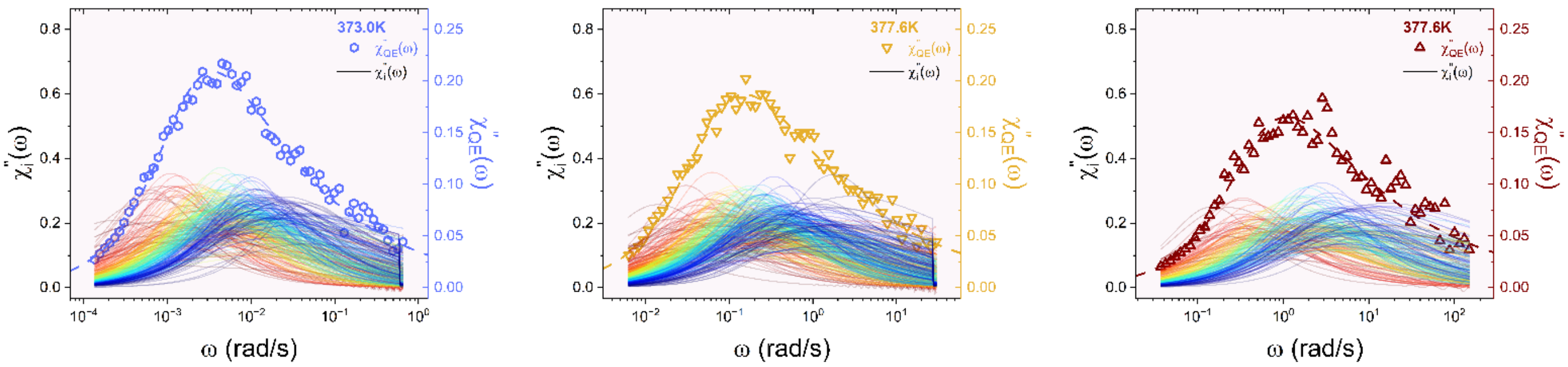

**Figure S3.** Thin colored curves show the dielectric-like loss spectra $\chi_i''(\omega)$ reconstructed for individual molecules from their time-domain rotational fluctuations using the Onsager-Kubo/FDT mapping (implemented via molecule-specific KWW fits; see Methods). Open symbols show the corresponding quasi-ensemble spectrum $\chi_{\mathrm{QE}}''(\omega)$ at the same temperature, obtained by applying the same reconstruction pipeline to the QE autocorrelation. $\chi_i''(\omega)$ are shown for the indicated temperatures (373.0 K, 377.6 K, and 380.9 K), illustrating substantial molecule-to-molecule variation in peak position and lineshape (dynamic heterogeneity) alongside the QE susceptibility.

## Supplementary 4 - Analysis Pipeline

## Supplementary Text

It is important to distinguish between the different ensemble-level susceptibilities that can be constructed from single-molecule data. The first is the QE susceptibility discussed throughout the manuscript, which is obtained by first constructing the QE autocorrelation function through direct averaging of the raw SM autocorrelation functions in the time domain, followed by Fourier transformation using the Onsager-Kubo/FDT mapping (an example of which is shown in Figure 4, green symbols). A second, mathematically equivalent, ensemble susceptibility can also be constructed by directly Fourier transforming each raw SM autocorrelation function individually and then averaging the resulting susceptibilities in the frequency domain. Because the Fourier transformation itself is a linear operation, this procedure yields the same numerical ensemble susceptibility as the QE susceptibility. However, while this approach preserves the ensemble-level information, the individual SM susceptibilities generated from the raw autocorrelation functions are noisy due to the limited signal-to-noise ratio of single-molecule trajectories, making molecule-resolved analysis difficult to interpret. Therefore, in this work, a different analysis pipeline for the SM data was adopted, in which each SM autocorrelation function was first fit to a stretched-exponential (KWW) form, and the corresponding susceptibility was reconstructed from the molecule-specific KWW fit parameters. These reconstructed SM susceptibilities were then averaged to generate another ensemble-level susceptibility, which exhibits a slightly different peak position and lineshape compared to the QE susceptibility (an example of which is shown in Figure 4, red curve).

Such differences originate not from the Fourier transformations, but from the manner in which the molecule-resolved susceptibilities are constructed. In the single-molecule analysis pipeline, each SM autocorrelation function is first represented by an individual KWW fit before transformation into the frequency domain. In practice, the least-squares KWW fitting was performed only over the portion of the autocorrelation decay above 0.1, as beyond this noise becomes increasingly dominant and the correlation function is no longer reliably resolved. Thus, the fitted KWW parameters are primarily determined by the early- and intermediate-time portions of the relaxation. In contrast, the QE susceptibility is obtained directly from the QE autocorrelation function, which is constructed by averaging the raw SM autocorrelation functions prior to Fourier transformation and therefore retains the full ensemble-averaged decay. As shown in Figure S4, progressively lowering the fitting cut-off from 0.3 to 0.001 systematically shifts the averaged KWW-reconstructed susceptibility toward the QE susceptibility, with both the spectral lineshape and peak position converging toward the QE result. This demonstrates that the observed discrepancy originates from the finite fitting range used in the molecule-resolved KWW reconstruction procedure rather than from the Fourier transformation itself. Technically, in the limit where the fitting cutoff approaches zero and the full autocorrelation decay is included in the KWW representation, the summed susceptibility reconstructed from the SM KWW fits will converge to the QE susceptibility obtained directly from the QE autocorrelation function.

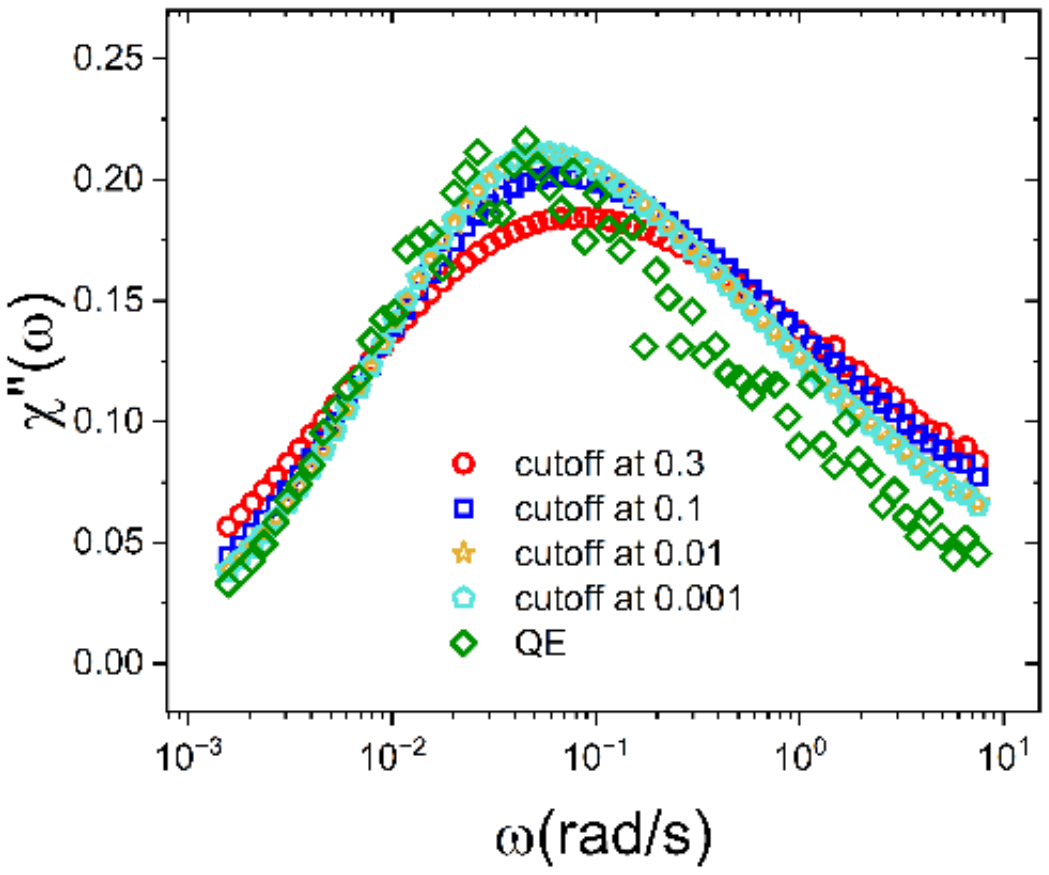

**Figure S4.** Effect of the KWW fitting cut-off on the constructed ensemble susceptibility spectrum at 375.8 K. Colored symbols (other than green) show the ensemble-averaged susceptibility obtained by averaging molecule-resolved loss spectra reconstructed from KWW fits performed over different fitting ranges of the SM autocorrelation functions. For each cut-off condition, the least-squares KWW fitting was performed only over the portion of the autocorrelation decay above the indicated correlation threshold before transformation into the frequency domain. Green diamonds are identical to the QE spectrum presented in Figure 4 in the main text and blue squares are identical to the red dotted line in Figure 4 in the main text.